\documentclass[twocolumn]{aastex631}
\usepackage{amsmath}
\usepackage{appendix}
\usepackage{float}
\usepackage{xcolor}
\usepackage{soul}
\usepackage[T1]{fontenc}
\usepackage{subfigure}
\usepackage{wrapfig}
\usepackage{hyperref}
\usepackage{chngpage}

\graphicspath{}
\DeclareUnicodeCharacter{2212}{-}

\begin{document}
\title{Variability Study and Searching for QPOs with day-like periods in the blazar S5~0716+714 with TESS}
\author{Shubham Kishore}
\affiliation{Aryabhatta Research Institute of Observational Sciences (ARIES), Manora Peak, Nainital, 263001, India}
\affiliation{Indian Institute of Astrophysics (IIA), Bangalore, 560034, India}
\author{Alok C.\ Gupta}
\affiliation{Aryabhatta Research Institute of Observational Sciences (ARIES), Manora Peak, Nainital, 263001, India}
\author{Paul J.\ Wiita}
\affiliation{Department of Physics, The College of New Jersey, 2000 Pennington Rd., Ewing, NJ 08628-0718, USA}
\email{amp700151@gmail.com (SK), acgupta30@gmail.com (ACG), wiitap@tcnj.edu (PJW)}
\begin{abstract}
\noindent
Using an unprecedented cadence of 30 minutes provided by the Transiting Exoplanet Survey Satellite (TESS), we have examined the optical light curves (LCs) of the blazar S5 0716+714 obtained from its Sectors 40, 47, and 53 over a period of about 75 days. This source exhibited flux variability in each of those sectors, reaching a maximum variability amplitude of 5.6\%. The power spectral density (PSD) shapes were tested with a simple power law and two distinct bending power laws and were found to be better fit by bending power laws than simple power laws for all but one of the segments. To look for any periodicities in these LCs, we used weighted wavelet Z (WWZ) transform analysis and generalized Lomb–Scargle periodograms (LSPs).  We identified one possible quasi-periodic oscillation (QPO) signature in a portion of sector 40 (period of $\sim$6.5~h), having $\sim$95\% global significance. A statistical approach to assess the light curves involving continuous autoregressive moving average (CARMA) was implemented, and the light curves were found to follow more complex processes than the simplest and typical damped random walk process. We briefly discuss the statistical properties of the light curves along with the general variability features and physical processes that could cause these types of fluctuations.
\end{abstract}
\section{Introduction}\label{sec:intro}
\noindent
Active galactic nuclei (AGNs) are extremely compact accretion-powered central regions of host galaxies involving astrophysical processes {produced by supermassive black holes (SMBHs) of $10^6-10^{10}\ M_\odot$} that are challenging to resolve even with the best available telescopes. 
Blazars, being a subclass of radio-loud and jetted AGNs pointing their jets almost along the observer's line of sight \citep[\(\lesssim 10^\circ\),][]{1995PASP..107..803U}, make up a very small fraction of all {AGNs ($<1\%$)}. It has been well established that blazars show flux, spectral, and polarization variations in all accessible electromagnetic (EM) bands on diverse timescales ranging as short as a few minutes to as long as several years \citep[e.g.,][and references therein]{1989Natur.337..627M,1993ApJ...411..614U,2001A&A...377..396R,2006A&A...459..731R,2012MNRAS.425.1357G,2016MNRAS.458.1127G,2022ApJS..260...39G}. Blazars are usually classified into two subclasses: BL Lacertae objects (BL Lacs) and flat spectrum radio quasars (FSRQs). BL Lacs show featureless  continua or very weak emission lines (i.e., equivalent width (EW) $\leq$ 5\AA) \citep{1991ApJS...76..813S,1996MNRAS.281..425M}, whereas FSRQs show prominent emission lines in the composite optical/ultra-violet spectra \citep{1978PhyS...17..265B, 1997A&A...327...61G}.\\
\\
Blazar flux variability across the entire EM spectrum is mostly non-linear, stochastic, and aperiodic \citep[e.g.,][]{2017ApJ...849..138K}, with emission being predominantly non-thermal \citep[e.g.,][and references therein]{2019AJ....157...95G}, and these flux changes provide an important way to probe these accretion-powered sources.
Very high flux variability in the form of flares, quiescent phases, and partially coherent {variations in the form of periodic or quasi-periodic oscillations} (QPOs) of various lengths are among the characteristic features of blazars \citep[e.g.,][and references therein]{2020NatCo..11.4176S, 2024ApJ...960...11K, 2021ApJS..253...10F}. No single AGN emission model has been able to explain all these observed traits, but these features can be employed to constrain some of the physical processes and accompanying mechanisms acting on short-length scales that are related to the timescales of variations \citep[e.g.,][and references therein]{2020NatCo..11.4176S, 2023ApJ...943...53K, 2024ApJ...968L..17V}.\\ 
\\
QPOs or periodic oscillations have been seen rather frequently in the {light curves (LCs)} of neutron star and stellar-mass BH binaries \citep{2006ARA&A..44...49R}. While the LCs of AGNs are primarily non-periodic throughout the whole EM spectrum, over the past two decades, sporadic QPO detections in several EM bands and with a range of periods have been documented in a number of blazars \citep[e.g.][and references therein]{2009ApJ...690..216G,2019MNRAS.484.5785G,2009A&A...506L..17L,2013MNRAS.436L.114K,2015ApJ...813L..41A,2016AJ....151...54S,2017A&A...600A.132S,2016ApJ...832...47B,2018NatCo...9.4599Z,2019MNRAS.487.3990B,2020A&A...642A.129S,2021MNRAS.501...50S,2020ApJ...896..134P,2024MNRAS.52710168P,2021MNRAS.501.5997T,2024ApJ...977..166T,2022MNRAS.510.3641R,2022MNRAS.513.5238R,2022Natur.609..265J,2023ApJ...943...53K,2023ApJ...950..173D} and other AGN subclasses \citep[e.g.][and references therein]{2008Natur.455..369G,2014MNRAS.445L..16A,2015MNRAS.449..467A,2016ApJ...819L..19P,2018A&A...616L...6G}. 
QPOs with timescales of days or hours are believed to be very rare in AGNs \citep[e.g.,][]{2018Galax...6....1G}, and their nature is unclear.  Multiple {possible} explanations for these features have been {suggested} \citep[e.g.,][]{2023ApJ...943...53K}, including
emission from plasma moving helically inside the jet \citep[e.g.,][]{2020A&A...642A.129S}, plasma instabilities \citep[e.g.,][]{2020MNRAS.494.1817D} or orbital
motion in an accretion disc \citep[e.g.,][]{2018AJ....155...31H,2022MNRAS.510.3641R}. \\
\\
The bright blazar S5~0716+714\footnote{\url{https://www.lsw.uni-heidelberg.de/projects/extragalactic/charts/0716+714.html}} ($z = 0.31\pm0.08$) \citep{2008A&A...487L..29N} has been among the most studied blazars, including examinations of variability across all EM bands on diverse timescales \citep[e.g.,][and references therein]{1990A&A...235L...1W,1996AJ....111.2187W,1991ApJ...372L..71Q,1997A&A...327...61G,2003A&A...400..477T,2003A&A...402..151R,2005A&A...429..427P,2005A&A...433..815B,2006A&A...455..871F,2008A&A...481L..79V,2012MNRAS.425.1357G,2013A&A...552A..11R,2013A&A...558A..92B,2015ApJ...809..130C,2016MNRAS.458.2350W,2017A&A...600A.132S,2018A&A...619A..45M}. Variations from S5~0716+714 are seen with a very high duty cycle {(a means to quantify the overall variability pattern, defined as the fraction of time that the source shows a variable nature based on available observational data)}  of $\sim$1 \citep[e.g.,][]{1995ARA&A..33..163W}, which means that the object is almost always in a somewhat active state. 
S5~0716+714 emits very high energy (VHE) $\gamma-$rays, as reported by {Major Atmospheric Gamma Imaging Cherenkov (MAGIC)} observations \citep{2009ApJ...704L.129A}, and has been included in the catalogue of TeV emitting sources\footnote{\url{http://tevcat.uchicago.edu/}}. Periodic oscillations or QPOs on diverse timescales have apparently been detected on several occasions in different EM bands from the blazar S5~0716+714 \citep[e.g.][]{1991ApJ...372L..71Q,1996A&A...305...42H,1996AJ....111.2187W,2003A&A...402..151R,2009ApJ...690..216G,2010ApJ...719L.153R,2013A&A...552A..11R,2017MNRAS.471.3036P}. It is possible that a potential QPO on a $\sim$1 day timescale was simultaneously detected in the radio and optical bands \citep{1991ApJ...372L..71Q}. On another occasion, its optical emission seemed to exhibit quasi-periodicity with a $\sim$4 days timescale \citep{1996A&A...305...42H}. In a multi-wavelength observational campaign of the source, a QPO with a period of a $\sim$7 days was detected at 5 GHz radio frequency \citep{1996AJ....111.2187W}. From a long-term optical LC, a QPO with a period of $3.0\pm0.3$ years was reported in this blazar \citep{2003A&A...402..151R}. In an extensive search for optical QPOs on the timescales of minutes using the published LC taken from 1999 to 2003 by \citet{2006A&A...451..435M}, \citet{2009ApJ...690..216G} found evidence for QPOs on 5 occasions in the period range of 25--37 minutes. \citet{2010ApJ...719L.153R} carried out optical observations of the source in 2008 and reported QPO detection on one occasion with the period of $\sim$15 minutes. In multi-wavelength observations of the blazar S5 0716+714 from 2008 to 2011, \citet{2013A&A...552A..11R} found evidence for a 63 days period in the optical band and \citet{2017MNRAS.471.3036P} suggested a 346 days period in $\gamma$-rays. \\
\\
{The mass of the central BH plays a crucial role in AGN physics, greatly influencing the accretion disc size, rate, and temperature via the Eddington limit. In addition, it also affects the relativistic jet power, variability timescales of disc instabilities, host galaxy’s bulge, position of the synchrotron, and inverse-Compton peak frequencies for blazars, etc., making it important to estimate the BH mass. 
There are several methods to assess the mass of the accreting BHs. The primary ones include stellar and gas kinematics investigation, requiring high-resolution host galaxy spectroscopy. Reverberation mapping that tracks the response of emission lines to continuum variations, and megamasers investigation \citep[e.g.,][]{1995PNAS...9211427M} are also the basic methods for BH mass deductions. However, all these methods rely on the presence of well-resolved spectral lines \citep[e.g.,][and references therein]{2000ApJ...540L..13P, 2002MNRAS.331..795M, 2003MNRAS.340..632L, 2005MNRAS.361..919P, 2013ApJ...773...90G}. Since the optical spectrum of the blazar S5 0716+714 is a featureless continuum, it is not possible to utilize the above-noted methods. An empirical method based on the BH mass and velocity dispersion of the host galaxy bulge is also inapplicable to the source, as the host galaxy of the source has only been marginally detected \citep{2008A&A...487L..29N}.} \\
\\
{In this paper, we present the time-domain statistical study of the optical LC of the blazar S5~0716+714, which was observed with the  Transiting Exoplanet Survey Satellite (TESS). This includes the LC's variability inspections and a thorough QPO search in the obtained LCs.} In Section 2, we describe data acquisition and reduction. In Section 3, we explain the data analysis techniques we used and present the results. A discussion is provided in Section 4.
\section{Data Acquisition and Reduction}
\label{section_2}
\begin{table}[h]
\caption{Flux calibration details}
\begin{adjustwidth}{-1cm}{}
\resizebox{3.7in}{!}{
\begin{tabular}{c c c c}\hline\hline
 Sector & $\alpha$ & Overfitting metric & Underfitting metric\\\hline
 40 & 0.1 & 0.803 & 0.990\\
 47 & 0.1 & 0.806 & 0.834\\
 53 & 0.1 & 0.841 & 0.961\\
\hline
\end{tabular}}
\end{adjustwidth}
\label{tab1:cal.details}
\end{table}
\begin{figure*}[t]
     \hspace{-1.15cm} 
     \includegraphics[width=1.1\linewidth]{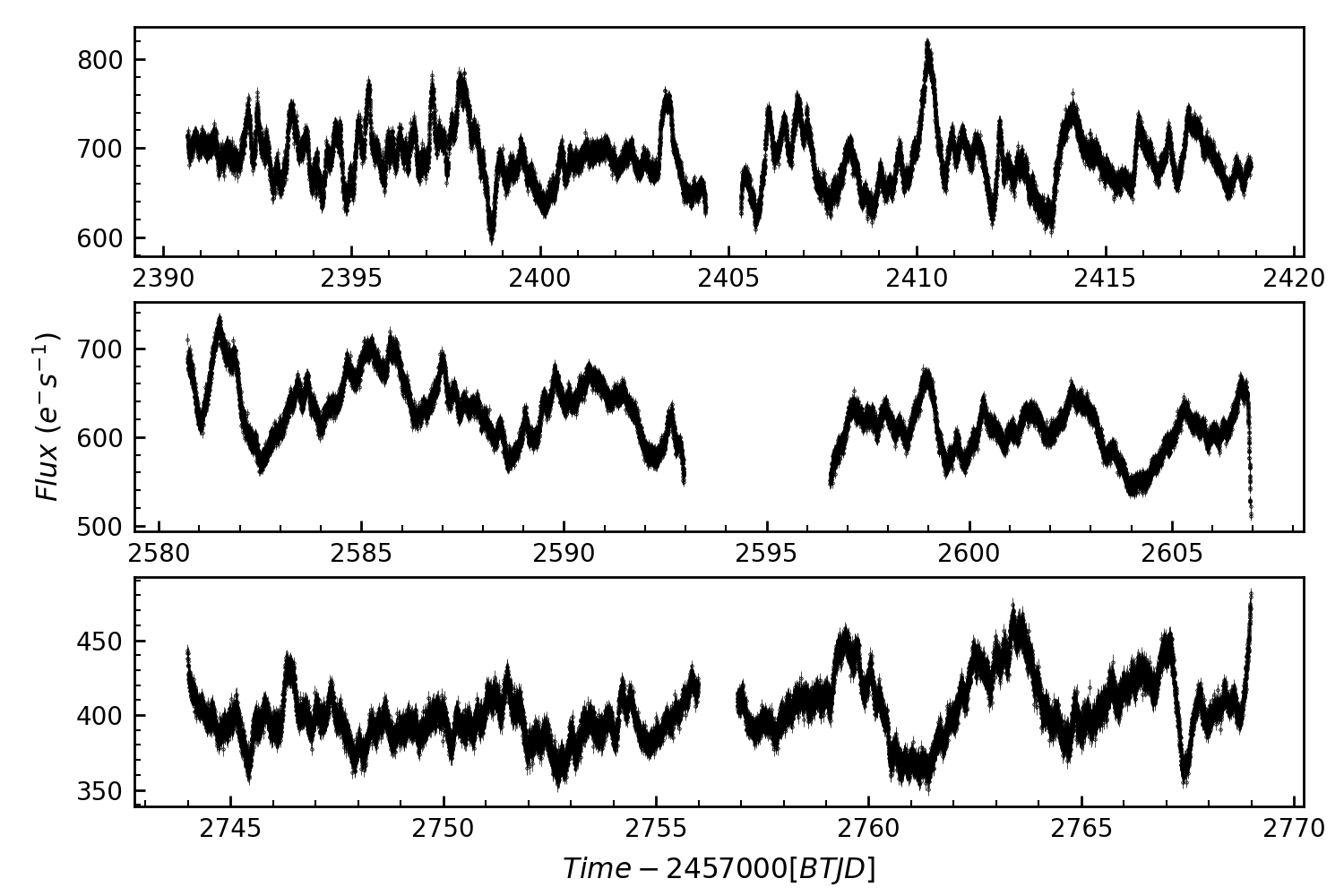}
     \caption{Light curves of S5 0716+714 corresponding to Sectors 40 (top), 47 (middle), and 53(bottom)}
     \label{fig:lightcurve}
\end{figure*}
\noindent Aside from some data gaps of 1 -- 2 days related to telemetry, S5 0716+714 was observed by TESS in eight Sectors: 20, 26, 40, 47, 53, 60, 73, and 74.  TESS's detector bandpass spans the range 600--1000 nm and is centered at the traditional Cousins I-band\footnote{\url{https://heasarc.gsfc.nasa.gov/docs/tess/the-tess-space-telescope.html}}. We obtained the optical SAP\_FLUX \citep[{flux obtained by summing constituent pixel values via simple aperture photometry, see}][{for details}]{2016SPIE.9913E..3EJ, 2017ksci.rept....9J} provided by the TESS Science Processing Operations Center (SPOC) pipeline for calibration with the co-trending basis vectors (CBVs) as described in \citet{2023ApJ...943...53K} and have followed the procedure given there. Hence, we refer the readers to that paper for the nomenclature used in this section and the optimum values found for the three parameters employed during data reduction: the two goodness metrics (overfitting and underfitting) and the regularization factor ($\alpha$). \\
\\ Unfortunately, there were no {\tt lightcurvefiles} already available for Sectors 60, 73, and 74 through the SPOC, and the overfitting metric values for Sectors 20 and 26 were found to be much lower than the minimum optimum value of 0.8 needed for proper calibration. Hence, we only considered data from Sectors 40, 47, and 53 in our inspection of the LCs (constituting a composite timespan of \(\sim\)74 days).
Table~\ref{tab1:cal.details} includes these parameters obtained for different sectoral reductions, and Fig.~\ref{fig:lightcurve} presents the LCs obtained for the three noted sectors after reduction. {In our analysis, each of the sectoral LCs was segmented into two parts, of roughly 12 days in length, separated by the standard data gap to ensure even spacing between the data points.}
\section{Data Analysis and Results}
\subsection{Excess variance} 
\noindent
As the LCs of blazars are degraded by instrumental noise, they can very often exhibit nominal variability signatures enhanced by that noise. Hence, the typical variance test can lead to erroneous results. Since flux variability acts as one of the fundamental properties of a typical blazar across all EM bands, incorporating the intrinsic measurement uncertainties becomes crucial.  The excess variance method helps take care of this uncertainty in estimating the intrinsic variability of the source. {It is estimated via}  
\begin{equation}
    F_{var}=\sqrt{\frac{S^2-\overline{\sigma_{err}^2}}{\bar{x}^2}},
\end{equation}
\begin{equation}
    S^2=\frac{1}{n-1}\sum(x-\bar{x})^2
\end{equation}
and
\begin{equation}
 (F_{var})_{err}=\sqrt{\left[\sqrt{\frac{1}{2n}}\frac{\overline{\sigma_{err}^2}}{F_{var}\bar{x}^2}\right]^2+\left[\sqrt{\frac{\overline{\sigma_{err}^2}}{n}}\frac{1}{\bar{x}}\right]^2},
\end{equation}
 where \(F_{var}\) is the fractional root mean square variability amplitude, \(S^2-\overline{\sigma_{err}^2}\) is the variance after discarding the average contribution expected from measurement uncertainties; \(\bar{x},\ \overline{\sigma_{err}^2}\ \text{and}\ n\) represent the mean flux value, mean square uncertainty and the number of flux data points.
The fractional root-mean-square variability amplitude, that is, the square root of the normalized excess variance, and the corresponding uncertainty, have been computed as in \citet{2003MNRAS.345.1271V}. We employed this excess variance as the first test to confirm the genuine variability in the LCs. Each of the six segments was tested individually with this, and the results have been presented in Table~\ref{tab2}. We found that the source shows a corrected variability amplitude of between 3.2\% and 5.6\% on these $\sim$12~d {intervals}. 
{\subsection{Periodogram Analysis}
\noindent
Blazar LCs often {exhibit} a `red noise' periodogram or power spectral density (PSD) where the periodogram shapes generally follow a simple power law relation $P(\nu)\sim\nu^{-\alpha}$ with $\nu$ the temporal frequency, so the power decreases rapidly with increasing frequency. {Aside from these simple power laws,  bending power laws or broken power laws} have also been observed to characterize the PSD shapes \citep[e.g.][]{2012A&A...544A..80G}. These characteristic structures and {PSD slopes can provide clues to} the states of the observed systems. An additional benefit of the periodogram is that any distinctive oscillatory behavior in the time series can be readily detected {and can be} tested for its significance against the corresponding PSD shape. }\\
\\
We have used the generalized Lomb-Scargle periodogram \citep[LSP;][]{1976Ap&SS..39..447L,1982ApJ...263..835S,2005MNRAS.361..919P, 2009A&A...496..577Z, 2018ApJS..236...16V} to do a segment-wise PSD analysis of the obtained LCs. 
The PSD shapes were fitted with three distinct PSD models \citep[e.g., see][]{2010MNRAS.402..307V,2013MNRAS.433..907E}:\\
$M_1$ (Simple Power law) \begin{equation}
    P(\nu)=A\nu^{-\alpha_1}+c
\end{equation}
$M_2$ (Bending Power law -- `A')
\begin{equation}
    P(\nu)=A\nu^{-1}\left[1+\left(\frac{\nu}{\nu_b}\right)^{\alpha_1-1}\right]^{-1}+c
\end{equation}
and $M_3$  (Bending Power law -- `B')
\begin{equation}
    P(\nu)=A\nu^{-\alpha_2}\left[1+\left(\frac{\nu}{\nu_b}\right)^{\alpha_1-\alpha_2}\right]^{-1}+c
\end{equation}
where {the free parameters} $A,~\alpha_1,~\alpha_2,~\nu_b~\text{and}~c$, are the normalization, spectral indices, bending frequency, and an additive constant, respectively. The above three models  {respectively have three, four, and} five free parameters. We have used the minimization of negative log-likelihood method to derive the free parameters for the above power law models. The function to be minimized {is given} as \citep[see][]{2010MNRAS.402..307V}
\begin{equation}
    \label{lglkd}
    \mathcal{L} = -2\sum_{j}\frac{I_j}{P_j}+\text{log }P_j
\end{equation}
where $I_j$ and $P_j$ are the observed periodogram and the model spectrum. The corresponding uncertainties on the model parameters were computed following \citet{2012A&A...544A..80G}. We further used the Bayesian Information Criterion (BIC) approaches to quantify the applicability of the individual models characterizing the PSD shapes, and to get the preferred model. This method relies on the negative log-likelihood minimization, additionally accounting for the number of free parameters, incorporating the number of data points as well. The preferred model has the lowest BIC value. It is given as\\
\begin{equation}
    BIC=k\cdot\text{ln}(n)~-~\mathcal{L}
\end{equation}
where $k$ and $n$ are the number of free parameters and the number of data points. The BIC values for the best-fit models are presented in Table~\ref{tab2}. We find that all the segments can be well described by $M_2$ except 1/47, which is better described by $M_1$.\\
\begin{figure*}[]
    \resizebox{1.\linewidth}{!}{\includegraphics[]{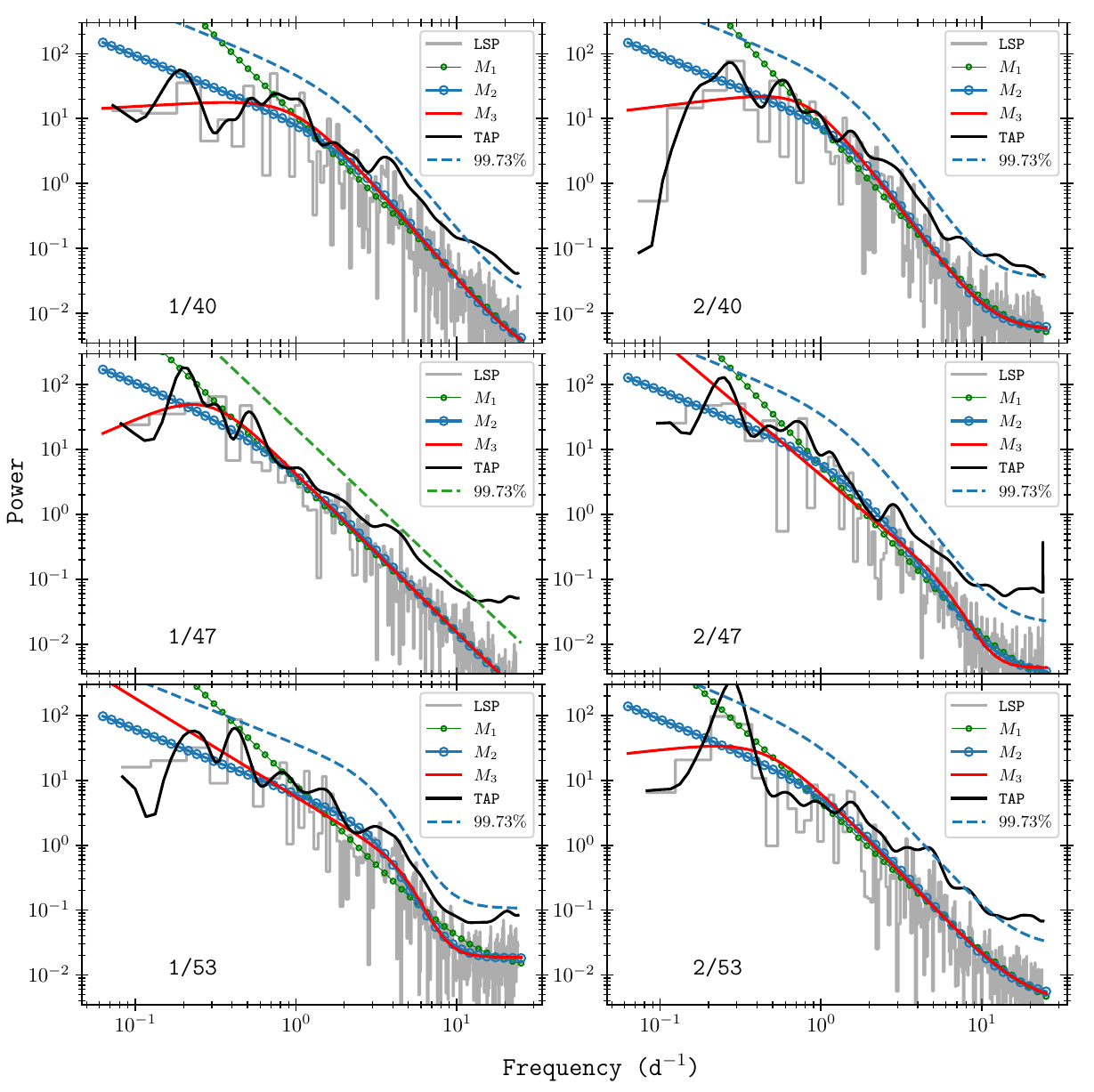}}
    \caption{LSPs and the TAPs obtained with the WWZ transforms for different indicated sectors. The three rows correspond to Sectors 40, 47, and 53, while in each row,  the two plots are for the two segments for each mentioned sector. The $99.73\%$ significance levels are the local ones and correspond to the most preferred power law model considered via the BIC coefficient.} 
    \label{fig:LSP}
\end{figure*}
\\
To estimate the distinct significance levels of any frequency peaks and the average spectrum for a comprehensive QPO search, we simulated \(10^5\) artificial LCs based on the mean, variance, PSD shape, and flux distribution of the LCs. To do this, we employed the publicly available Python code {\tt DELCgen\footnote{\href{https://github.com/samconnolly/DELightcurveSimulation/tree/master}{DELCgen.py}}}, a version of the LC simulation algorithm from \cite{2013MNRAS.433..907E},  and the LSPs of each individual created LC were estimated. However, with this method, the average LSPs of several segments could not properly fit the periodograms (Apparently there were slight disagreement in periodogram shapes estimated with the LSP and {\tt DELCgen.py} package for each individual segment: an example plot for the segment 1/40 depicting the poor fitting is presented in the Appendix (Fig.~\ref{fig:applots} -- Upper panel)). So we followed a rather straightforward method that uses the property that the periodogram is distributed in a \(\chi^2\) distribution with two degrees of freedom \citep[Eqns.~15 and 16 of][]{2005A&A...431..391V}. Then we can estimate a proper multiplication factor for the required significance levels. The best fits with the three models are presented in Fig.~\ref{fig:LSP} and the LSP fit parameters are included in Table~\ref{tab2}. Using appropriate multiplication factors, Fig.~\ref{fig:LSP} includes the local $99.73\%~(3\sigma)$ significance levels for corresponding best models for each of the segments. We find that none of the segments as a whole give any strong periodicity peak crossing the local $3\sigma$ significance level, making it unnecessary to compute global levels, as that would further increase the threshold levels.
\begin{table*}
\caption{Variability amplitudes, LSP fit parameters for simple power and bending power law fits, and corresponding BIC values; the uncertainties in $F_{var}$ values are less than $10^{-4}\%$ and hence not presented.\\
\noindent $\dagger$: The simple power law fitting of LSP of this segment offered a negative value of c, so we employed two approaches: one includes fixing it to zero, and the other, when it was set to the value obtained with the model $M_2$.}
\hspace{-2.5cm}
\resizebox{20.5cm}{!}{\begin{tabular}{c c c c c c c c c c}\hline\hline
Sector/ & timespan& $F_{var} $ & Normalization & Bending freq.& \multicolumn{2}{c}{Power spectral index}& Offset & BIC & CARMA\\
Segment & (d) & (\%) & constant (A)&(\(\nu_b\): (day\(^{-1}\))) & $\alpha_1$ & $\alpha_2$ & (c) & coef. & $(p,q)$\\
\hline\\
40/1$^\dagger$ & 13.73 & 4.06  &$(1.08_{-0.06}^{+0.06})\times10^1$ & & $2.47_{-0.02}^{+0.02}$ & & 0  & -1325.3 & (3,1)\\
&&& $(1.27_{-0.08}^{+0.08})\times10^{1}$ & & $2.61_{-0.03}^{+0.03}$ & & $2.38\times10^{-3}$ &  -1315.8\\
&&& $9.40_{-0.59}^{+0.63}$ & $2.13_{-0.07}^{+0.08}$ & $3.15_{-0.04}^{+0.05}$ & & $(2.38_{-0.54}^{+0.55})\times10^{-3}$ & -1344.9\\
&&& $(2.17_{-0.13}^{+0.13})\times10^1$ & $1.01_{-0.02}^{+0.02}$ & $2.82_{-0.03}^{+0.03}$ & $(-1.48_{-2.32}^{+1.18})\times10^{-1}$ &$(1.37_{-0.53}^{+0.53})\times10^{-3}$ &  -1344.6\\\\
40/2 & 13.52 & 4.70  &  $7.46_{-0.57}^{+0.60}$ & & $2.76_{-0.04}^{+0.04}$ & & $(4.18_{-0.50}^{+0.50})\times10^{-3}$ & -1593.3 & (3,1)\\
&&& $9.32_{-0.81}^{+0.91}$ & $1.62_{-0.07}^{+0.07}$ & $3.61_{-0.08}^{+0.08}$ & & $(5.89_{-0.52}^{+0.53})\times10^{-3}$ & -1619.9\\
&&& $(3.08_{-0.26}^{+0.28})\times10^1$ & $(8.42_{-0.22}^{+0.22})\times10^{-1}$ & $3.26_{-0.05}^{+0.05}$& $(-2.98_{-2.37}^{+2.39})\times10^{-1}$ & $(5.44_{-0.52}^{+0.52})\times10^{-3}$ & -1617.7\\\\
47/1$^\dagger$ & 12.23 & 5.38 & $3.55_{-0.20}^{+0.21}$ & & $2.36_{-0.02}^{+0.03}$ & & 0 & -1687.0 & (3,2)\\
&&& $3.85_{-0.24}^{+0.25}$ & & $2.44_{-0.03}^{+0.03}$ & & $6.1\times10^{-4}$ & -1684.9 & \\
&&& $(1.10_{-0.07}^{+0.07})\times10^{1}$ & $(7.04_{-0.29}^{+0.28})\times10^{-1}$ & $2.63_{-0.02}^{+0.03}$ & & $(6.1_{-2.7}^{+2.6})\times10^{-4}$ & -1682.9\\
&&& $(3.55_{-0.22}^{+0.21})\times10^2$& $(2.86_{-0.05}^{+0.05})\times10^{-1}$ & $2.45_{-0.02}^{+0.02}$ & $-1.08_{-0.05}^{+0.05}$& $(2.3_{-1.8}^{+2.6})\times10^{-4}$ &-1679.0\\\\
47/2 & 10.38 & 4.42 & $5.56_{-0.44}^{+0.48}$ & & $2.68_{-0.04}^{+0.04}$ & & $(2.51_{-0.39}^{+0.40})\times10^{-3}$ & -1357.7 & (3,1)\\
&&& $8.07_{-0.72}^{+0.79}$ & $1.50_{-0.07}^{+0.07}$ & $3.39_{-0.07}^{+0.07}$ & & $(3.46_{-0.40}^{+0.41})\times10^{-3}$ & -1370.6\\
&&& $4.01_{-0.39}^{+0.44}$ & $7.15_{-0.43}^{+0.48}$ & $6.18_{-0.68}^{+0.82}$ & $2.09_{-0.07}^{+0.07}$ &$(4.33_{-0.40}^{+0.41})\times10^{-3}$ & -1367.8\\\\
53/1 & 11.98 & 3.21  & $8.33_{-0.69}^{+0.74}$ & & $2.47_{-0.05}^{+0.05}$ & & $(1.24_{-0.14}^{+0.14})\times10^{-2}$ & -954.8 & (3,0)\\
&&& $6.13_{-0.62}^{+0.69}$ & $3.26_{-0.14}^{+0.14}$ & $4.74_{-0.22}^{+0.24}$ & & $(1.81_{-0.14}^{+0.15})\times10^{-2}$ & -987.9\\
&&& $5.53_{-0.57}^{+0.64}$ & $5.07_{-0.24}^{+0.26}$ & $6.42_{-0.60}^{+0.70}$ & $1.52_{-0.09}^{+0.09}$ & $(1.89_{-0.14}^{+0.15})\times10^{-2}$ &-985.8\\\\
53/2 & 12.06 & 5.60  & $4.54_{-0.30}^{+0.32}$ & & $2.32_{-0.03}^{+0.03}$ & & $(2.14_{-0.57}^{+0.59})\times10^{-3}$ & -1361.3 & (3,1)\\
&&& $8.74_{-0.69}^{+0.74}$ & $1.23_{-0.06}^{+0.06}$ & $2.89_{-0.05}^{+0.05}$ & & $(4.43_{-0.60}^{+0.61})\times10^{-3}$ & -1367.3\\
&&& $(5.71_{-0.42}^{+0.45})\times10^{1}$ &$(4.78_{-0.13}^{+0.13})\times10^{-1}$ & $2.58_{-0.03}^{+0.03}$ & $(-2.84_{-1.02}^{+0.99})\times10^{-1}$  & $(3.48_{-0.60}^{+0.59})\times10^{-3}$ & -1361.9\\\\
\hline \\
\end{tabular}}\hfill
\label{tab2}
\end{table*}
\subsection{Weighted Wavelet Z (WWZ) Analysis}
\begin{figure*}[t]
    \vspace{-.6cm}
    \centering
    \includegraphics[width=1.\linewidth]{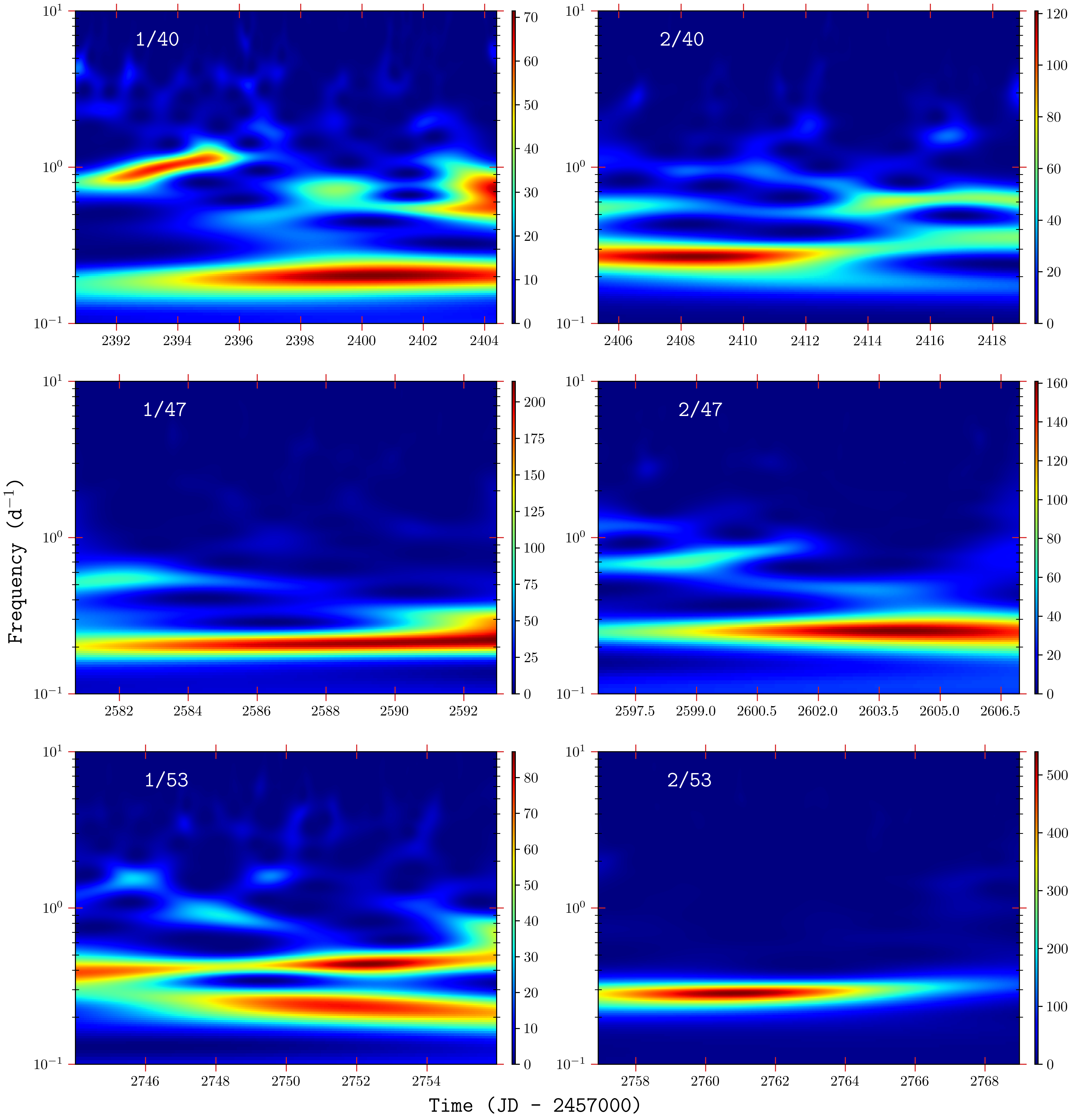}
    \caption{WWZ analysis of the segmented light curve. The three rows correspond to Sectors 40, 47, and 53, while in each row,  the two plots are for the two segments for each mentioned sector.} 
    \label{fig:wwz}
\end{figure*}
\noindent
The WWZ provides a particularly useful way to analyze LCs that might possess transient QPOs. When considering a limited temporal span of time-series data, this technique simultaneously decomposes these data into both frequency and time domains.  Thus, the WWZ analysis method can assess the temporal evolution of prominent periodicities, along with their phase and amplitude. Characterized by two basic properties, scale and position, a wavelet represents a localized wavelike oscillation. The transform utilizes these wavelets instead of the sinusoidal components, and they act as window/weight functions to constrain the portions of interest in the LC \citep[see][and references therein]{1996AJ....112.1709F}, to offer a two-dimensional map of power as a function of time and frequency.\\
\\
{Following \cite{2023ApJ...943...53K}, we performed the segmentwise WWZ analysis of the observed LCs to get a general overview of the behavior of the LC in the time-frequency domain.}  To do so we used the {\tt libwwz} python package\footnote{https://github.com/ISLA-UH/libwwz}, based on {the} \cite{1996AJ....112.1709F} and \cite{2004JAVSO..32...41T} Fortran codes. This way, we can make a visual inspection of LC behaviors without being immediately concerned about the significance of any periodic peculiarities. The WWZ maps of all the segmented LCs are presented in Fig.~\ref{fig:wwz}. \\
\\
We require that a genuine QPO signal must satisfy the criteria that the LSP peak corresponding to that signal must cross the local 99.73\% significance level above the most preferred underlying spectrum/model (selected through the BIC criteria), and covers at least \(\sim\)4 cycles (visualized via the WWZ map). An additional benefit of the WWZ analysis is that a time-averaged periodogram (TAP: included in Fig.~\ref{fig:LSP}), or mean of the WWZ, can also be constructed from the WWZ map by integrating it along the time axis; this is similar to the nominal power spectral density of the LC and helpful in detecting periodicities.
Every visually indicated high-powered feature in the WWZ map was considered a region of interest and was tested for periodicity searches. One way in which disagreement among different techniques may arise is from a difference in considering the associated uncertainties with the individual data points. The LSP plots in all segments contain several peaky structures. Since any sustained signal in the WWZ maps naturally produces a corresponding peak or hump-like structure in the TAP, we only consider LSP features supported by a corresponding distinctive TAP peak/hump as possible signals of QPOs.
\subsection{Stochastic Modeling}
\noindent The PSDs of LCs can also be estimated using autoregressive (AR) methods. One such method includes {\tt REDFIT} analysis \citep[][an established FORTRAN package]{2002CG.....28..421S} that is based on fitting the LCs using the simplest AR(1) model. We tested our LCs with the {\tt REDFIT} program but found that this approach could not properly fit the periodogram, particularly in the higher frequency regions (An example plot of periodogram fitting for the segment 1/40 showing the poor fitting using {\tt REDFIT} method is presented in the Appendix (Fig.~\ref{fig:applots} -- lower panel). So, we employed a more complex technique: CARMA (Continuous AutoRegressive Moving Average), to assess the segmented LCs. This approach combines continuous forms of autoregressive (AR) and moving average (MA) models under the assumption that the LCs are outcomes of a Gaussian process $y(t)$, given via stochastic differential equation \citep[see][and references therein]{2014ApJ...788...33K} as:
\begin{equation}
\begin{aligned}
\frac{d^py(t)}{dt^p} + &\alpha_{p-1}\frac{d^{p-1}y(t)}{dt^{p-1}}+\cdots + \alpha_0y(t)\\
=&\;\beta_q\frac{d^q\epsilon(t)}{dt^q}+\beta_{q-1}\frac{d^{q-1}\epsilon(t)}{dt^{q-1}}+\cdots + \epsilon(t)
\end{aligned}
\end{equation}
where $\epsilon(t)$ ia a continuous white noise process and $(\alpha_i,\beta_i)$ are the autoregressive and moving average coefficients. Basically, this model relates the LC and its first p-derivatives with a noise process and its first q-derivatives, which for stationary processes requires $p>q$, and provides more flexibility in fitting both LCs and the shapes of the PSDs of the LCs \citep[see][]{2017MNRAS.470.3027K}. In other words, CARMA models deal with the past memories of the stochastic processes and associated random fluctuations on different timescales. The {\tt REDFIT} algorithm is equivalent to the CARMA~(1,0) model, which has been found to be insufficient to  describe the segmented LCs. \\
 \\
 Blazar LCs are well known to be stochastic in nature, and  often cannot be fit by the simplest CARMA~(1,0), also known as the damped random walk  process.    Higher order CARMA models and associated modified versions have been observed to better fit the random nature of AGN LCs across multiple wavelength regimes of the EM spectrum \citep[e.g., see][and references therein]{2017MNRAS.470.3027K, 2018ApJ...863..175G, 2020ApJS..250....1T, 2022ApJ...930..157Z, 2024ApJ...960...11K, 2024A&A...690A.223K}. We have utilized the publicly available Python code {\tt EzTao} following \citet[][]{Yu2022} to find the simplest possible CARMA models which describe the segmented LC behaviours. We consider all pairs $(p,q):~1<p<5,~q<p$
for the CARMA fittings. The best CARMA model $(p,q)$ was estimated by again minimizing the negative log-likelihood values, determined for all sets of $(p,q)$ with the above conditions \citep[see][for details]{2017AJ....154..220F}. The best CARMA fit $(p,q)$ pair values have been included in Table~\ref{tab2}. We find that all the segmented LCs require a minimum of $p=3$.
\section{Discussion}
\noindent
Here we have performed a segment-wise temporal examination of the TESS LCs of the blazar S5 0716+714 that were observed in three Sectors: 40, 47, and 53. 
This source was also observed in Sectors 20, 26, 60, 73, and 74, but for Sectors 20 and 26, no optimal reduction was attainable with the CBV detrending, and for Sectors 60, 73, and 74, no SAP\_FLUX was available with the SPOC pipeline.
We used the 30-minute binned LCs of the source in our analysis and first estimated the amount of variability in the six segmented LCs. For PSD characterization, we employed the traditional Lomb-Scargle periodograms approach, and for the periodicity searches, we also performed WWZ analyses to assess the LC in the time-frequency domain. In all these Sectors, the WWZ maps of the obtained LCs reveal some high-power regions, which were further tested with the same LSP. Simultaneous agreement of both of these techniques, along with the corresponding LSP peak touching or crossing a local 99.73\% significance level, and a minimum of 4 cycles, was considered a signature of a genuine QPO. \\
\\
During the complete observation, the source underwent several phases of variation, showing both intraday and short-term variabilities. The segment-wise analysis of the blazar S5 0716+714 indicated possible signatures of {dominant variations} in the range (\(\sim 0.4 - 1\ \text{d}^{-1}\)), in most of the segments with one at \(\sim\)0.58~d\(^{-1}\) during segment 40/2, but the LSPs analysis could not separate them from the background bending power law spectrum (see Fig. \ref{fig:LSP}). Of all the six segments, only one feature in the segment~1/40 might possibly be called a QPO feature. An independent analysis of a constrained portion of segment 1/40 reveals that the LSP feature at around $3-4~$d$^{-1}$ (corresponding to a period of around 6.5~hrs) in this segment only approaches a global significance of $\sim95\%$. Apart from this, segment 1/53 also shows some variations on similar timescales, but it has even lower global significance. These variations have relatively low powers (because of the red noise nature of the blazar LC), and due to their short-lived nature, further searches for and investigation of such signatures would be desirable. 
 It is also worth noting that the normalized power of the higher frequency peaks may be greatly diminished because of the relative increase of powers at lower frequencies if there are long-term trends.
\subsection{Light curves and periodogram properties} 
\noindent
The sectoral LCs of the source (Fig.~\ref{fig:lightcurve}) illustrate S5 0716+714 in three somewhat different flux states. All the LCs have been found to follow bending power law spectra, characterized by model $M_2$ except that of segment 1/47, which has a preference for model $M_1$. The quantity \(\alpha_1\) acts as a proxy for the spectral slope after the bending frequency, and the source has displayed an appreciably elevated range of slopes, untypical for blazars, which generally lie in the vicinity of 2. Except for the segment 1/47, the bending frequencies are found to be in $1.2 - 3.3$ range. In the segment 1/53, the best fitting spectral slope is quite steep at  \(\sim\)4.74. The bending frequency for this segment (1/53) is also high compared to those of the other bending power law fit segments, meaning that the micro-variations are highly dominant over day-timescales. It is noteworthy that the source happens to be in the lowest of the states in Sector 53. It is possible that these differences might be attributed to the accretion disk processes, as the contribution of the accretion discs can only be perceived when the jet contribution is minimum in the blazars (i.e., in a low flux state). \\
 \\
 For background periodogram estimation we also tested the LCs fitting with the AR(1) model using the {\tt REDFIT} programs \citep[see][]{2002CG.....28..421S} which is efficient in estimating red noise power spectra of time series data, but the theoretical (analytical) spectrum obtained from fitting the LCs could not properly fit the power spectrum of the LCs,  specifically in the high-frequency regime. This suggests that they require higher-order regression models such as CARMA ($p,q$), which are able to produce more complex LC behaviors, and they resulted in a preferred CARMA-$p$ value of 3.
 \subsection{Possible causes of aperiodic erratic variability}
 \noindent
 {Fig.~\ref{fig:lightcurve} reveals the existence of several short-term and intraday random variations displayed by the source. These variations are characteristic features of the AGNs though the the distinctive process causing such variations are highly debated, and several models have been put forward in this regard. Since S5~0716+714 is a blazar source, the dominant variations in the LC can be safely presumed to arise within the jets. The random variations highly support the idea of existence of ensemble of emitting regions/blobs. Geometrical origins involving motion of these blobs along a helical magnetic field can  describe the observed variability through changes in the Doppler factor, produced by small variations in our viewing angles to the individual blobs \citep[e.g.][]{2015ApJ...805...91M}. The change in Doppler factor does not change the intrinsic flux of the source but can significantly affect the observed flux along the observer. The turbulent extreme
multizone (TEMZ) model put forward by \cite{2014ApJ...780...87M} is another model capable of  producing this type of rapid variability. In this scenario, hot plasma blobs with randomly oriented magnetic fields flow along the jet, and when they 
cross a standing conical shock they can produce fast flux variations. This mechanism is also able to cause rotation of polarization angles. 
Another promising class of models that can yield such rapid variability are those involving ``jets-in-a-jet" or ``mini-jets" as proposed by \cite{2008MNRAS.386L..28G} and \cite{2009MNRAS.395L..29G}. In these scenarios the plasma in small regions of the jet is accelerated to very high Lorentz factors ($> 100$) via magnetocentrifugal or magnetic reconnection processes.}
\section*{ACKNOWLEDGMENTS}
\begin{figure*}[h]       
    \centering
    \includegraphics[width=.49\linewidth, trim=.15cm 0 1.25cm 0, clip]{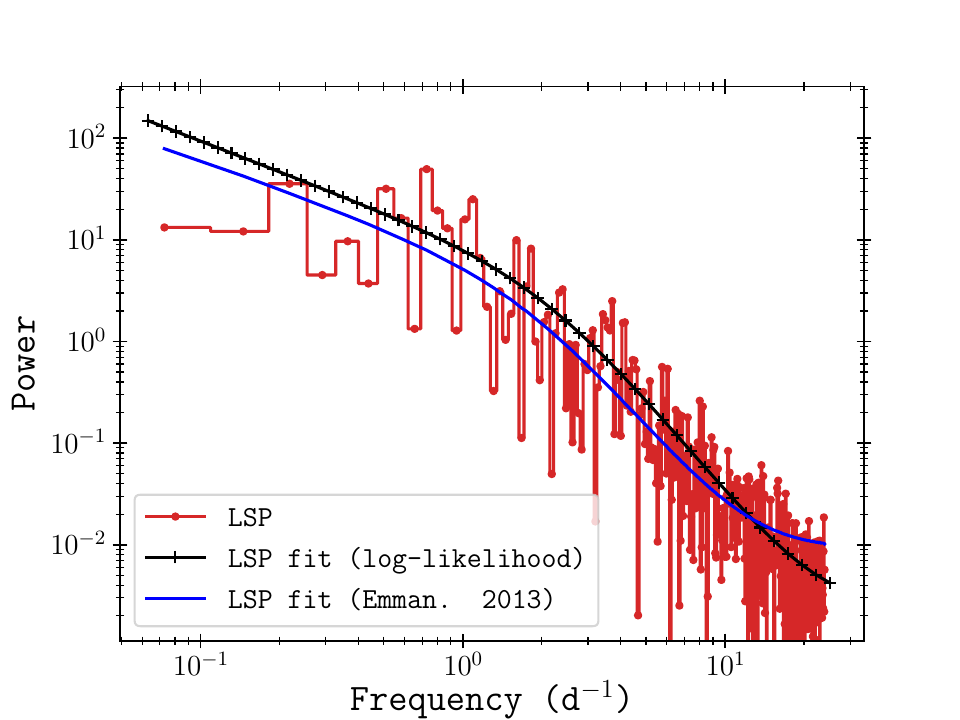}
    \includegraphics[width=.49\linewidth, trim=.15cm 0 1.25cm 0, clip]{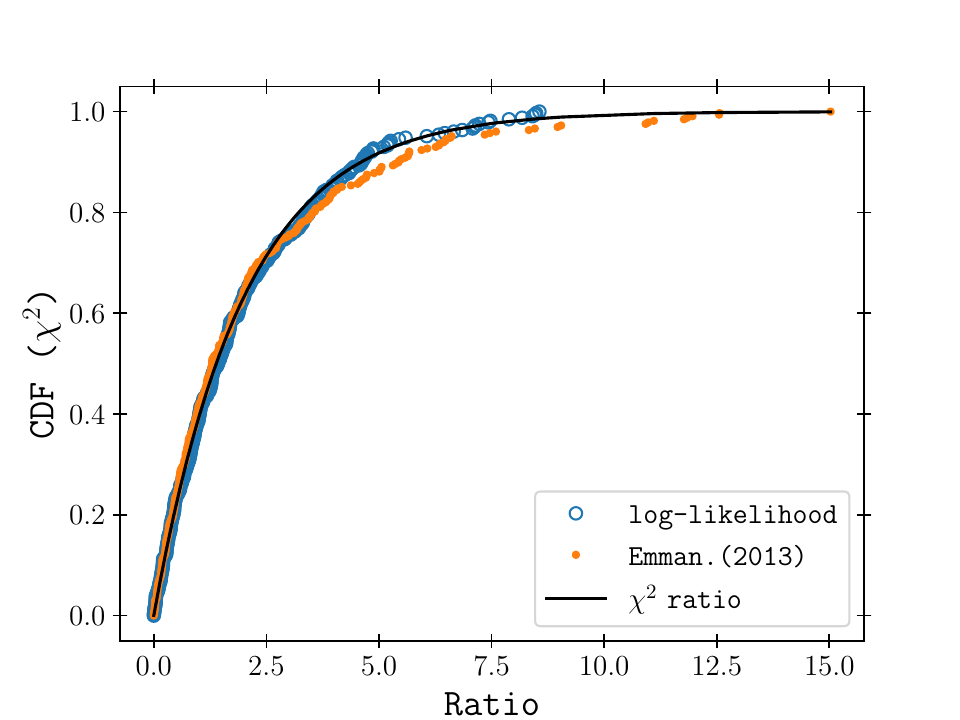}
    \includegraphics[width=.5\linewidth]{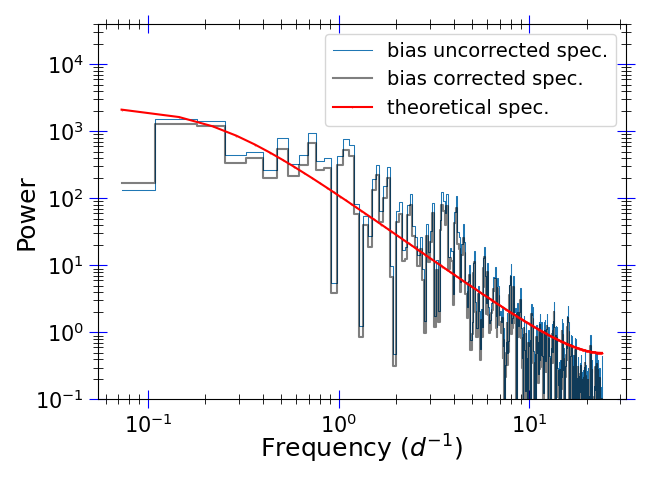}
    \caption{Upper left panel: Comparision of LSP fittings with the log-likelihood method and another by simulating $10^5$ lightcurves with {\tt DELCgen.py}; Upper right panel: Comparison of $\chi^2$ ratio CDF with the theoretical distribution of two degrees of freedom; Lower panel: PSD fitting via {\tt REDFIT} fit}
    \label{fig:applots}
    \end{figure*}

\noindent
 This paper includes data collected with the TESS mission, obtained from the MAST data archive at the Space Telescope Science Institute (STScI). Funding for the TESS mission is provided by the NASA Explorer Program. STScI is operated by the Association of Universities for Research in Astronomy, Inc., under NASA contract NAS 5–26555. \\
\\
{\it Facility:}  Transiting Exoplanet Survey Satellite (TESS) -- The dataset used in this paper can be found in MAST: \dataset[10.17909/32y4-kj47]{http://dx.doi.org/10.17909/32y4-kj47}. \\
\\
{\it Software:} lightkurve \citep{2018ascl.soft12013L}, 
SciPy \citep{2020SciPy-NMeth}, PyAstronomy \citep{pya}.\\
\bibliography{ref} 
\bibliographystyle{aasjournal}
 \appendix
 \noindent
 We present here (Fig.~\ref{fig:applots}) an example comparison of periodogram fittings for the segmant 1/40 using two methods. One follows \cite{2013MNRAS.433..907E} with which we obtained an average periodogram from the LSPs of $10^5$ simulated light curves based on flux distribution of input light curve and its PSD shape determined with the module from {\tt DELCgen.py}. The second directly fits the LSP of the input light curve via a log-likelihood method. The goodness of fit is evaluated by comparing the corresponding cumulative distribution function (CDF) with that of the theoretical $\chi^2$ distribution of two degrees of freedom \citep[see][]{2005A&A...431..391V}. We implemented the standard nonparametric Kolmogorov-Smirnov (KS) test, and the latter approach has been found to offer a better p-value under the null hypothesis that the two fits describe the obtained LSP. We also show here that the simplest stochastic approach of AR(1) with the {\tt REDFIT} method gives a visibly poor fit to the periodogram for the same segment, implying that the light curve is better fit using higher order regressive processes than the simplest AR(1).
\end{document}